\documentclass[final,5p,times,twocolumn,numbers,sort&compress]{elsarticle}

\usepackage{amsmath}
\usepackage{amssymb}
\usepackage{color}
\usepackage{hyperref}

\usepackage{lipsum}

\newcommand{\smaf}[2] {{\textstyle \frac{#1}{#2} }}

\journal{Nuclear Physics B}

\begin{document}

\begin{frontmatter}



\title{The LHC has ruled out Supersymmetry -- really?}


\author[a]{L. Constantin}
\author[a]{S. Kraml}
\author[b,c,d]{F. Mahmoudi}

\affiliation[a]{organization={Laboratoire de Physique Subatomique et de Cosmologie (LPSC), Université Grenoble-Alpes, CNRS/IN2P3}, 
            postcode={F-38026}, 
            city={Grenoble},
            country={France}}

\affiliation[b]{organization={Université Claude Bernard Lyon 1, CNRS/IN2P3, Institut de Physique des 2 Infinis de Lyon, UMR 5822},
            postcode={F-69622}, 
            city={Villeurbanne},
            country={France}}

\affiliation[c]{organization={Theoretical Physics Department, CERN},
            addressline={CH-1211 Geneva 23}, 
            country={Switzerland}}

\affiliation[d]{organization={Institut Universitaire de France (IUF)},
            postcode={F-75005}, 
            city={Paris},
            country={France}}

\begin{abstract}
Despite early hopes that the Large Hadron Collider (LHC) would quickly unveil supersymmetric particles, none have been detected to date. This review examines the impact of the LHC results on the viability of weak-scale supersymmetry, and discusses whether the possibility of discovering supersymmetric particles remains within reach. 
\end{abstract}


\end{frontmatter}




\section{Introduction}
\label{introduction}

Supersymmetry (SUSY) is a theoretical framework that extends the Standard Model (SM) by a new symmetry between bosons and fermions. In this scenario, every known particle (more precisely, every SM degree of freedom) has a supersymmetric partner,\footnote{Here, we consider SUSY with $\mathcal{N}=1$ generator. For $\mathcal{N}>1$, supermultiplets grow larger.} differing in spin by half a unit. 
SUSY has been the prime paradigm for new physics beyond the SM for decades. For a detailed introduction, see e.g.\ \cite{Martin:1997ns}.

One of the key motivations is the so-called hierarchy problem: within the SM, quantum corrections drive the Higgs boson mass to extremely high values, requiring a fine-tuning to maintain it at the electroweak scale. Weak-scale SUSY naturally solves this issue by ensuring that the quadratically divergent corrections from fermionic and bosonic partners cancel out, thereby stabilising the Higgs mass. 
Moreover, through  renormalization group running, the large top Yukawa coupling can drive the mass-squared term of (one of) the Higgs doublet scalars negative near the electroweak scale, thus naturally triggering  electroweak symmetry breaking.
Additionally, SUSY predicts the unification of gauge couplings at high energy scales. Another compelling aspect of supersymmetric models is the ability to provide viable dark matter candidates, particularly in models where R-parity is conserved: in this case, the lightest supersymmetric particle (LSP), often assumed to be a neutralino, remains stable and could account for the observed dark matter density in the universe.

If R-parity is conserved, SUSY particle production at colliders is characterised by a distinctive experimental signature: SM particles accompanied by significant missing transverse energy (MET). The LSP, which remains stable under these conditions, naturally fulfills the role of a weakly interacting massive particle (WIMP) and can account for the observed dark matter abundance over a broad parameter space.

Early expectations for the discovery of SUSY at the LHC were high, particularly for strongly interacting superpartners. Theoretical arguments, but also global fits of constrained SUSY scenarios ---such as the  Constrained Mininal Supersymmetric Standard Model (CMSSM)--- incorporating constraints from flavour physics, low-energy precision measurements, and dark matter relic density (see e.g.~\cite{Allanach:2007qk,Buchmueller:2008qe,Trotta:2008bp}), suggested that superparticle masses should be within reach of the LHC initial runs, typically below or around 1~TeV. However, searches by the \mbox{ATLAS} and CMS experiments (for a review, see e.g.\ \cite{Canepa:2019hph}) have ruled out substantial regions of this parameter space, challenging initial expectations.\footnote{Comprehensive overviews of search results, including summary plots, are  available at the ATLAS and CMS wiki pages \url{https://twiki.cern.ch/twiki/bin/view/AtlasPublic/SearchesPublicResults} and \url{https://cms-results.web.cern.ch/cms-results/public-results/publications/SUS/index.html}}

Indeed, mass limits are being pushed higher and higher by the LHC experiments. The current bounds reported by ATLAS and CMS reach up to about $2.4$~TeV for gluinos and squarks of the first two generations, $1.3$~TeV for stops and sbottoms,  and $1$~TeV for electroweak-inos~\cite{SUSYatMoriond25}. 
It has to be kept in mind, however, that these bounds are derived in the context of simplified models, assuming the presence of only few (typically 2--3) new particles and simple decay patterns (often a single decay channel). 
Realistic SUSY models can be much more complicated, with a large variety of relevant states and intricate mixing patterns affecting cross sections, decay branching ratios, and in specific cases even the lifetimes of the new states. Thus, even in the simplest realistic model, the so-called Mininal Supersymmetric Standard Model (MSSM), mass bounds can be weaker than those listed above.

Be as it may, the absence of conclusive evidence for supersymmetry at the LHC has led to growing frustration within the high-energy physics community. 
Many researchers nowadays question whether SUSY remains a viable solution to the shortcomings of the SM, and, in popular discourse, there is a growing perception that the LHC has effectively ``ruled out" SUSY. 

In this article, we address the question whether this claim is true or an oversimplification.

\section{Higgs mass, hierarchy and naturalness problems}

\begin{figure}[t]\centering
    \includegraphics[width=0.48\textwidth]{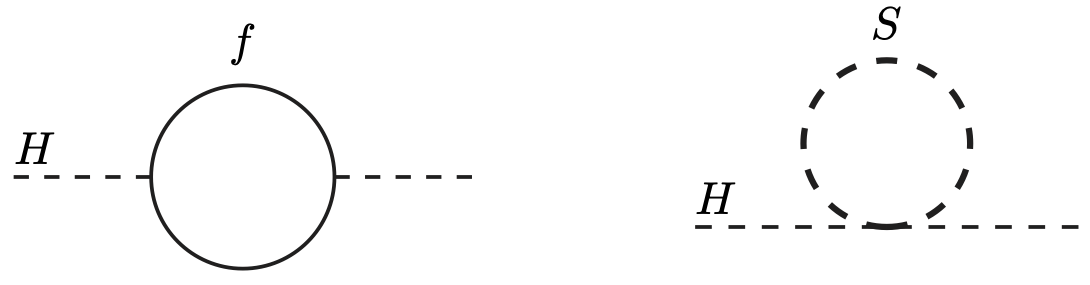}
    \caption{One-loop corrections to $m_H^2$, due to a Dirac fermion $f$ (left), and a scalar $S$ (right); from \cite{Martin:1997ns}.
\label{fig:higgscorr}}
\end{figure}

In order to break the electroweak symmetry and give masses to the 
$W$ and $Z$ bosons and the matter fermions, 
some scalar field must acquire a non-zero vacuum expectation value (VEV). 
In the SM, this field is elementary, leading to an elementary scalar 
``Higgs'' boson of mass $m_H$. However, elementary scalar masses are 
notoriously unstable in the SM. 
For example, the one-loop diagram shown in Fig.~\ref{fig:higgscorr}(left) 
is quadratically divergent, yielding  
\begin{equation}
  \delta m_H^2  = {\cal O}\left({\alpha \over 4 \pi}\right) \,\Lambda^2 \,,
  \label{eq:quaddiv}
\end{equation}
where the cutoff $\Lambda$ represents the scale up to which the SM remains valid; 
this may be $M_X\ge 10^{15}$~GeV in a grand unified theory (GUT), or 
the Planck scale $M_{\rm P}\sim 10^{19}$~GeV, if no new physics sets in before. 
Moreover, in GUTs, the couplings of a light Higgs boson to the heavy Higgs 
fields present in the theory also yield $\delta m_H^2  = {\cal O}(M_X^2)$. 
This is in principle not a problem in a renormalisable theory. 
However, it requires an extreme fine-tuning of counterterms order by order in pertubation theory, and this seems unnatural.\footnote{For an in-depth discussion of the hierarchy problem, we refer to the article by M.E.~Peskin~\cite{Peskin:2025lsg} in this volume.} 

These large corrections to the SM Higgs boson mass, which should be 
$m_H={\cal O}(m_W)$, raise problems at two levels:\\ 
(i) to arrange, in the first place, for $m_H$ to be many orders smaller
than other fundamental mass scales such as $M_X$ or $M_{\rm P}$ 
---the hierarchy problem, 
and (ii) to avoid corrections $\delta m_H^2$ to the Higgs boson mass squared,  
which are much larger than $m_H^2$ itself ---the naturalness problem.

In SUSY, SM particles and 
their superpartners have the same couplings but differ in spin by half a unit. 
Since fermion (Fig.~\ref{fig:higgscorr} left) and boson (Fig.~\ref{fig:higgscorr} right) loop diagrams 
have opposite signs, the quadratic sensitivity to the cutoff $\Lambda$ 
in eq.~(\ref{eq:quaddiv}) cancels. The remaining correction is  
\begin{equation}
\delta m_H^2 = {\cal O}\left( \frac{\alpha}{4\pi} \right) \left(m_S^2 - m_f^2 \right) \log\left( \frac{\Lambda^2}{Q^2} \right)\, ,
  \label{eq:quaddiv1}
\end{equation}
where $m_f$ and $m_S$ stand for the masses of the SM particles and their 
superpartners, and $Q$ is typically of the order of SUSY-breaking scale.
Thus, a residual logarithmic sensitivity to the ultraviolet cutoff remains due to soft SUSY breaking. 
This is no larger than the physical value, $\delta m_H^2 \lesssim m_H^2$, provided  
\begin{equation}
|m_S^2 - m_f^2| \lesssim \mathcal{O}(1~\text{TeV})^2 \cdot \left[\log\left( \frac{\Lambda^2}{Q^2} \right)\right]^{-1} \,.
\end{equation}
This is (or was) the main motivation to expect SUSY to be realised at the TeV energy scale! 
The argument is in fact most severe for stops ($\tilde t$), the scalar partners of the top quark, since the largest contribution to $\delta m_H^2$ in the SM comes from top-quark loops. Therefore the fact that LHC searches seem to push $m_{\tilde t}\gtrsim 1$~TeV is perceived as a severe problem. 

On the other hand, in SUSY, the Higgs boson mass is not a free parameter like in the SM. In the MSSM, the Higgs sector consists of two complex doublets $H_u=(H^+_u,H^0_u)^T$, $H_d=(H^0_d,H^-_d)^T$ with the scalar potential for the (neutral) Higgs fields given by
\begin{align}
  {\cal V} = 
   &\:(|\mu|^2+m_{H_u}^2) \, |H^0_u|^2
   +  (|\mu|^2+m_{H_d}^2) \, |H^0_d|^2 
   - (b H_u^0H_d^0 + {\text h.c.}) \notag\\[1mm]
   & + \smaf{1}{8}(g_1^2+g_2^2)\,(|H_u^0|^2 - |H_d^0|^2)^2 
\,.  
\label{eq:scalhiggspot}
\end{align}
Here $\mu$ is a SUSY-conserving Higgs mixing parameter from the superpotential (${\cal W}\in \mu H_uH_d$),  while $m_{H_u}^2$, $m_{H_d}^2$ and $b$ are 
soft SUSY-breaking parameters; see \cite{Martin:1997ns} for details.
The structure is that of a two-Higgs doublet model of type~II and leads to five physical Higgs states, a neutral CP-odd, two neutral CP-even, and two charged Higgs bosons denoted by $A^0$, $h^0$, $H^0$, and $H^\pm$, respectively. 
An important point here is that the quartic term in eq.~\eqref{eq:scalhiggspot} is a function of the gauge couplings, in contrast to being a free parameter in the SM. As a consequence, after electroweak symmetry breaking, there is an upper bound on the mass of the lighter of the two CP-even Higgs bosons $h^0$: 
\begin{equation}
    m_{h} \leq m_Z\, |\cos 2\beta| 
\end{equation}
at tree level. 
Here, $\tan\beta=v_u/v_d$ is the ratio of the VEVs at the minimum of the potential, $v_u\equiv\langle H_u^0\rangle$ and 
$v_d\equiv\langle H_d^0\rangle$, and $v^2\equiv v_u^2+v_d^2 = 2\,m_Z^2/(g_1^2+g_2^2)$.

\begin{figure}[t!]\centering
    \includegraphics[width=0.4\textwidth]{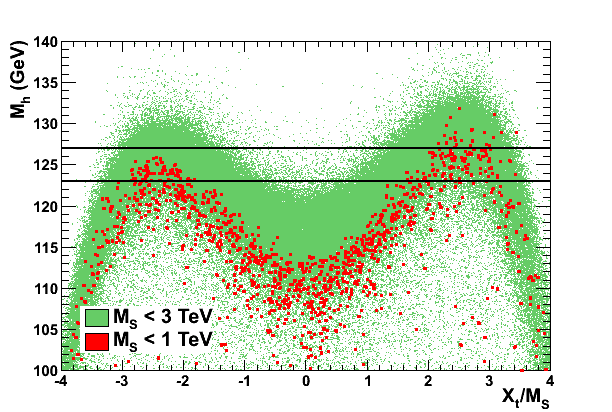}
    \caption{Dependence of the light MSSM Higgs boson mass, $M_h$, on the ratio $X_t/M_S$; from \cite{Arbey:2011ab}.}
    \label{fig:maxmix}
\end{figure}

Fortunately, $m_{h}$ is subject to large radiative corrections. Again, due to the large top Yukawa coupling, the largest effect comes from the top-stop sector.  
In the decoupling limit $m_{A}\gg m_Z$, one gets at the one-loop level 
\begin{equation}\label{eq:mh02loop}
  m_{h}^2 = m_Z^2\cos^2 2\beta 
  + \frac{3}{4\pi^2}\frac{m_t^4}{v^2}
  \left(\log\frac{M_S^2}{m_t^2}+\frac{X_t^2}{M_S^2}\left(1-\frac{X_t^2}{12\,M_S^2}\right)\right)\,.
\end{equation}
Here $M_S^2=m_{\tilde t_1}m_{\tilde t_2}$ with $m_{\tilde t_{1,2}}$ the stop masses, and $X_t=A_t-\mu\cot\beta$ is the stop mixing parameter. To achieve the observed value of $m_{h}=125$~GeV \cite{ATLAS:2012yve,CMS:2012qbp} either the average stop mass must be large, $M_S\equiv\sqrt{m_{\tilde t_1}m_{\tilde t_2}} \gtrsim 3$ TeV, or the stop mixing parameter $|X_t|$ must be around twice $M_S$, as illustrated in Fig.~\ref{fig:maxmix} (taken from \cite{Arbey:2011ab}, who, together with \cite{Brummer:2012ns}, also discuss the implications for GUT models).\footnote{There is a vast literature on this subject, and on the topic of precision calculations of $m_{h}$. For reviews, we refer the reader to \cite{Djouadi:2005gj,Draper:2016pys}.} This leads to a tension with the expectation of light stops (and not too heavy gluinos, due to their loop contributions to stop masses)
motivated by the hierarchy problem, and with the wish for small radiative corrections to satisfy naturalness arguments. 

Models beyond the MSSM can alleviate the problem through extra tree-level contributions to $m_{h}^2$. This is the case, for instance, in models with extra singlet scalars\ \cite{Ellwanger:2009dp}, or when gauginos are Dirac instead of Majorana states \cite{Benakli:2012cy}. 
Either way a tension remains and has led many people to doubt SUSY as a solution to the hierarchy and naturalness problems. 

One way out is to abandon the hierarchy problem altogether and use gauge-coupling unification and dark matter as the only guiding principles, as proposed in \cite{Arkani-Hamed:2004ymt,Giudice:2004tc,Arkani-Hamed:2004zhs}.  
In this case, all the MSSM scalars become (ultra)heavy, except for a single finely tuned SM-like Higgs boson. Yet, the fermionic superpartners (gauginos and higgsinos) can remain light, protected by chiral symmetry, and account for the successful unification of gauge couplings. This is known as Split SUSY and has distinctive consequences for LHC signatures. 

An even more extreme approach is to consider that gauginos and higgsinos are also very heavy, with masses near the SUSY-breaking scale $M_S$. This defines the High-Scale SUSY framework~\cite{Hall:2009nd}, where supersymmetry is effectively decoupled from the low-energy spectrum. In this scenario, one forgoes not only the naturalness argument but also the supersymmetric explanation for dark matter---since the LSP is no longer light---and the successful gauge coupling unification typically offered by SUSY.
Nevertheless, supersymmetry leaves an imprint at low energies through the matching conditions between the high-scale SUSY theory and the effective Standard Model. In particular, the Higgs quartic coupling $\lambda$ retains information about the underlying SUSY structure. As a result, even when SUSY is broken at very high energies, it still predicts a relatively light Higgs boson, with its mass carrying indirect sensitivity to parameters like $M_S$ and $\tan\beta$.

\begin{figure}[t!]\centering
    \includegraphics[width=0.4\textwidth]{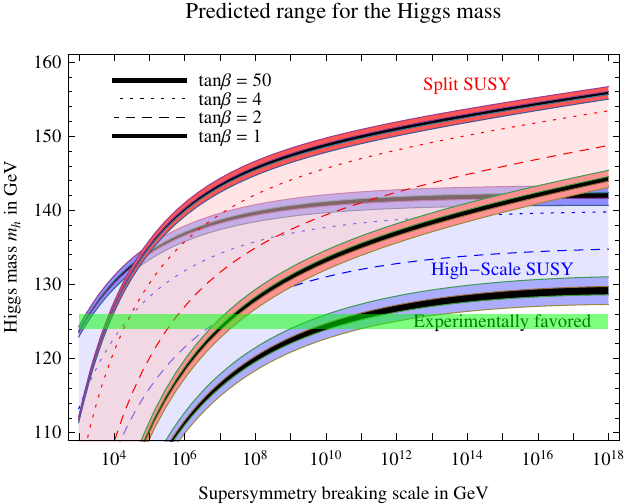}
    \caption{Predictions for the Higgs mass $m_h$ in Split SUSY (red) and in High-Scale SUSY (blue) for $\tan\beta=\{1,2,4,50\}$; from \cite{Giudice:2011cg}.}
    \label{fig:splitsusy}
\end{figure}

The relation between the observed Higgs mass at 125~GeV and the SUSY mass scale in the Split SUSY and High-Scale SUSY scenarios was investigated in \cite{Giudice:2011cg,Arbey:2011ab,Bagnaschi:2014rsa}
and is illustrated in Fig.~\ref{fig:splitsusy}.\footnote{This may be strongly influenced by tree-level contributions to $\lambda$ from a non-minimal field content.}

\section{Little Hierarchy problem and electroweak naturalness}

One can also adopt the more optimistic point of view of Baer et al., that SUSY offers a \mbox{'t Hooft} technically natural solution to the (Big) Hierarchy problem, and what matters is the electroweak finetuning. This is related to the 
Little Hierarchy problem:
why is there a gap between the weak scale $m_{weak}\sim m_{W,Z,h}\sim 100$ GeV
and the apparently lowest-possible soft SUSY breaking scale $M_S > 1-2$ TeV?

In the MSSM, the weak scale is related to the soft SUSY breaking terms and the $\mu$ parameter by the electroweak minimisation condition:
\begin{equation}
\frac{m_Z^2}{2} = \frac{(m_{H_d}^2+\Sigma_d^d)-(m_{H_u}^2+\Sigma_u^u)\tan^2\beta}{(\tan^2\beta -1)}
-\mu^2\simeq -m_{H_u}^2-\mu^2 .
\label{eq:mzs}
\end{equation}
where the $\Sigma_{u,d}^{u,d}$ contain a large assortment of SUSY loop
corrections.
Barbieri and Giudice~\cite{Barbieri:1987fn} introduced the naturalness measure 
\begin{equation}
    \Delta_{\rm BG}\equiv max_i \left| \frac{\partial\log m_Z^2}{\partial\log p_i} \right| \,,
\end{equation}
where the $p_i$ are fundamental parameters of the theory, and used this to set bounds on SUSY masses by imposing a maximally allowed value of $\Delta_{\rm BG}^{\rm max}=10$. Following this idea, Ref.~\cite{Baer:2015rja} formulated a naturalness principle stating that 
``An observable ${\cal O}$ is natural if all {\it independent} contributions to ${\cal O}$
are less than or of order ${\cal O}$''. This is dubbed ``practical naturalness''.
Applied to eq.~\eqref{eq:mzs}, practical naturalness requires each term on the right-hand-side (RHS) of the equation to be 
$\lesssim m_Z^2/2$. 
This is used to define an electroweak fine-tuning measure $\Delta_{\rm EW}$ as 
\begin{equation}
    \Delta_{\rm EW}=max\left|\, {\rm{each~term~on~RHS~of~eq.~\eqref{eq:mzs}}}\, \right| /(m_Z^2/2) \,.
\label{eq:dew}
\end{equation}
Note that $\Delta_{\rm BG}$ agrees with $\Delta_{\rm EW}$ for both, weak-scale and high-scale inputs, when evaluated in terms of genuinely independent  parameters. 
Moreover, the naturalness problem related to $m_h^2$ evoked in the previous sections becomes much less severe in terms of practical naturalness. 

The authors of \cite{Baer:2015rja} conclude that gluinos, top squarks and other SUSY particles can live at the multi-TeV scale with
little cost to naturalness (because they have loop-suppressed
contributions to the weak scale), but higgsinos, whose masses are essentially given by $\mu$,
must lie close to the weak scale, with $\mu \lesssim 200-400$~GeV. 
As we will see below, such light higgsinos have a tendency to escape detection at the LHC. 
Even at the High-Luminosity LHC, much but not all of the $\mu \lesssim 200-400$~GeV parameter space can be covered, see \cite{Baer:2020sgm}.

\begin{figure}[t!]\centering
    \includegraphics[width=0.5\textwidth]{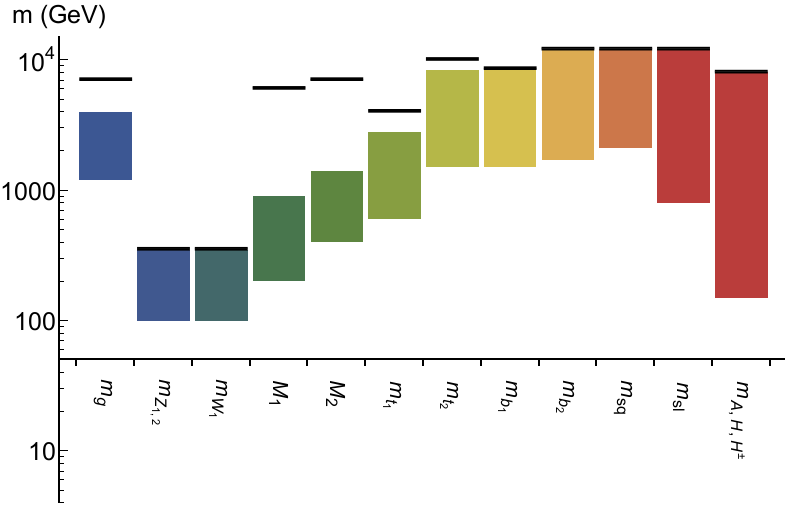}
    \caption{Range of sparticle and heavy Higgs  masses allowed by naturalness within the NUHM2 model with $\Delta_{\rm EW}<30$. Also imposed are the LEP2 limit on charginos, LHC Run~1 bounds on gluino and squark masses, and $m_h=125\pm2$~GeV. The black bars show upper bounds from the pMSSM model with 19 weak scale parameters; from \cite{Baer:2015rja}.}
    \label{fig:naturalsusy}
\end{figure}

Figure~\ref{fig:naturalsusy} shows the mass ranges compatible with $\Delta_{\rm EW}<30$ in the Non-Universal Higgs Mass Model (NUHM2). 
Also shown are the upper bounds withing the  
phenomenological MSSM (pMSSM), which is the topic of the next section.

\section{Phenomenological MSSM}
\label{sec:pmssm}

The pMSSM is a 19-dimensional realisation of the R-parity conserving MSSM with
parameters defined at the SUSY scale $M_S=\sqrt{m_{\tilde t_1}m_{\tilde t_2}}$. 
It was introduced in \cite{MSSMWorkingGroup:1998fiq} to provide a more general framework for studying MSSM phenomenology than the predominant minimal or constrained SUSY-breaking models, without having to deal with the $>100$ free parameters of the general MSSM, which would be unwieldy for most phenomenological purposes. 

\begin{figure}[t!]\centering
    \includegraphics[width=0.36\textwidth]{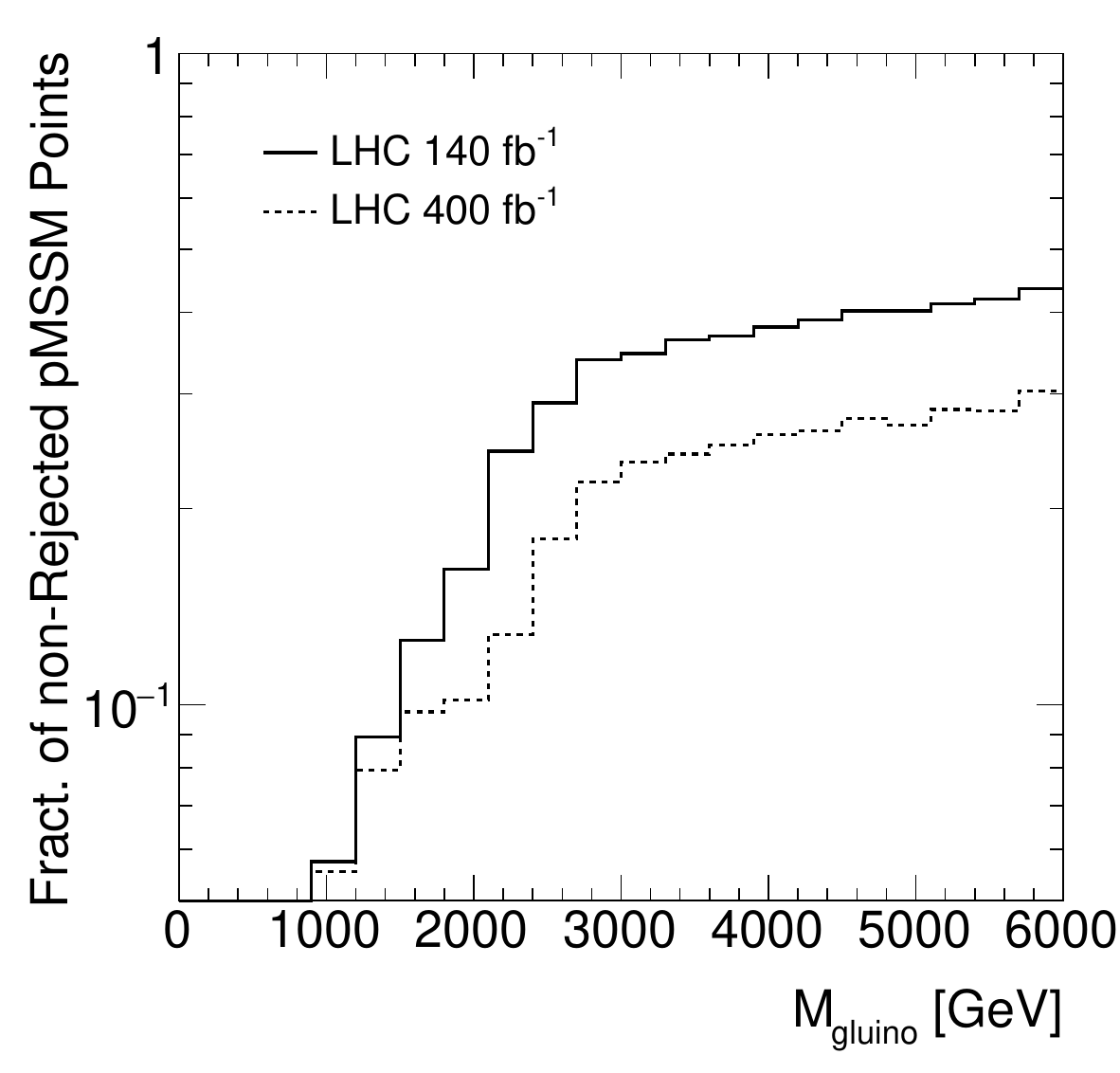} \\        \includegraphics[width=0.36\textwidth]{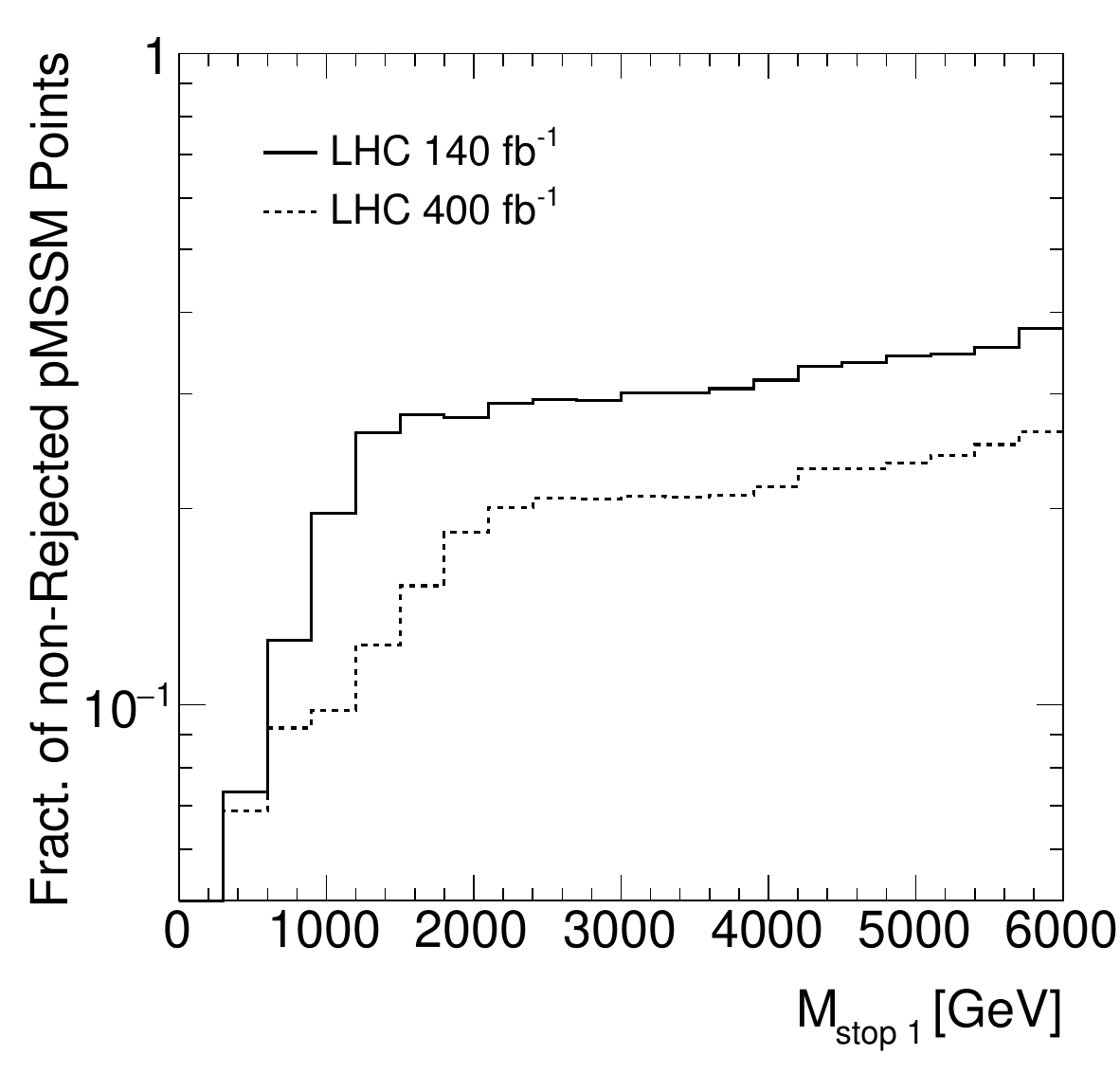}\\
\caption{Fraction of pMSSM points that are not excluded at 95\% confidence level by ATLAS SUSY searches from Run~2 (given in Table 4 of \cite{Arbey:2021jdh}), and that remain beyond the projected sensitivity of Run~3, shown as a function of the gluino mass (upper panel) and the lighter stop mass (lower panel); from \cite{Arbey:2021jdh}.}
\label{fig:msqlhc}
\end{figure}

The pMSSM employs only a few plausible 
assumptions motivated by experiment. 
Specifically, all soft SUSY-breaking parameters are taken to be real to avoid introducing new CP-violating effects. The sfermion mass matrices and trilinear coupling matrices are assumed to be diagonal, preventing FCNCs at tree level. Additionally, the first and second generations of sfermions are treated as degenerate at low energy, and trilinear couplings are assumed to be the same for all three generations. This leaves the pMSSM with 19 free parameters.

This setup allows for generic mass and mixing patterns, leading to a much larger variety of cross sections, decay channels and branching ratios, and thus to a much broader set of predictions for experimental signatures than those found in any of the constrained scenarios, such as the CMSSM, mSUGRA\footnote{minimal SUperGRAvity (mSUGRA) is a scenario in which SUSY breaking is mediated by gravity.}, NUHM models, etc., which impose quite restrictive relations between the soft SUSY-breaking parameters 
(see \cite{AbdusSalam:2009qd,Berger:2008cq,Conley:2010du,Sekmen:2011cz,Arbey:2012ntc} for pioneering studies).
Likewise, it also leads to a much broader set of predictions than realised in simplified models.

Consequently, a significant portion of the pMSSM parameter space escapes the bounds from ATLAS and CMS searches, which were set within constrained or simplified models. This was  confirmed by the ATLAS and CMS reinterpretations of their Run~1 searches \cite{ATLAS:2015wrn,CMS:2016lcl}.
Indeed, mass limits derived within simplified models are often overly optimistic and do not 
hold in the more general case of the pMSSM (or other realistic SUSY models).  

As an example, Fig.~\ref{fig:msqlhc} shows the fraction of pMSSM points (after flavour, dark matter, and low-energy constraints obtained using the SuperIso program~\cite{Mahmoudi:2007vz,Mahmoudi:2008tp}) not excluded by ATLAS Run 2 searches as a function of the gluino and stop masses. In the context of simplified models, the considered ATLAS analyses (given in Table 4 of \cite{Arbey:2021jdh}) exclude gluino masses up to $2.2-2.35$~TeV depending on the LSP mass, and 
 stop masses in the range $450-950$~GeV for 
LSP masses below 160 GeV (in the case where $m_{\tilde t_1}\sim m_t+m_{\tilde\chi^0_1}$, stops in the range of $235-590$ GeV are excluded).
This is to be compared with the full lines in Fig.~\ref{fig:msqlhc}, which indicate that gluinos with masses as low as 1 TeV and stops as light as 400 GeV remain viable within the pMSSM framework.\footnote{It should be noted that the figure is not intended to provide a statistical interpretation; rather, it serves to illustrate that viable spectra with relatively light superpartner masses are indeed allowed within the pMSSM framework.} 
Also shown in Fig.~\ref{fig:msqlhc} are the 
projected sensitivities of Run~3 with 400/fb (dashed lines). As one can see, the fraction of surviving points is reduced, but the overall conclusion remains unchanged. 

\begin{figure}[t]\centering
\includegraphics[width=0.48\textwidth]{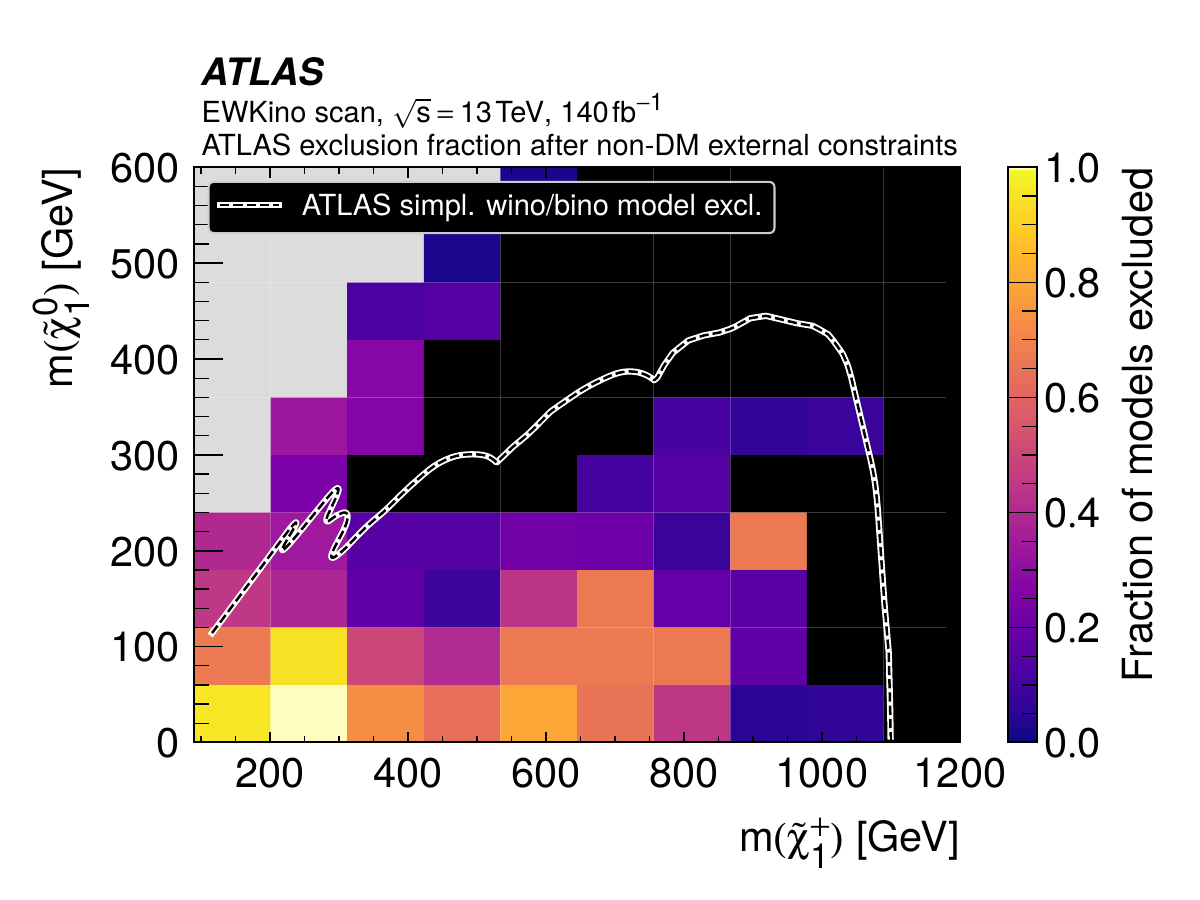}\caption{\small Fraction of electroweak-ino scenarios excluded by \mbox{ATLAS} Run~2 results. Considered are 7956 pMSSM points that have a neutralino LSP, satisfy chargino mass constraint from LEP and have a light Higgs mass in the range $m_h=120-130$~GeV. The white line shows the limit in the simplified wino/bino model; from \cite{ATLAS:2024qmx}.}
\label{fig:atlasewk}
\end{figure}

Turning to electroweak-inos (i.e.\ charginos and neutralinos, mixtures of electroweak gauginos and higgsinos), in \cite{ATLAS:2024qmx} the ATLAS collaboration provided an interpretation of their Run~2 searches for electroweak production of supersymmetric particles within the pMSSM. To this end, they scanned bino, wino and higgsino mass parameters $M_1$, $M_2$ and $\mu$ from $-2$~TeV to $2$~TeV; the gluino mass parameter 
$M_3=1-5$~TeV; 
$m_A=0-5$~TeV and $\tan\beta=1-60$. The masses of 1st/2nd generation sfermions were fixed to 10~TeV and the 3rd generation parameters scanned over to obtain scenarios with a SM-like Higgs in the right mass range. 
Considering the results from eight separate searches for electroweak-inos (see Table~5 in \cite{ATLAS:2024qmx}), each using 140/fb of proton-proton data from Run~2, it was found that 
many pMSSM scenarios with chargino and neutralino masses well below the simplified-model limits escape exclusion, as exemplified in Fig.~\ref{fig:atlasewk}.  

Apparently, in the pMSSM, the total SUSY signal tends to spread  out over several experimental channels. It is therefore interesting to ask how the sensitivity improves when statistically combining the likelihoods from different, independent  analyses. Likewise, it is interesting to combine ATLAS and CMS results, to increase statistics and mitigate fluctuations in the data. This is possible using the SModelS framework as demonstrated recently in  \cite{MahdiAltakach:2023bdn,Altakach:2023tsd}. 
Since \cite{ATLAS:2024qmx} made all their scan points publicly available, we can use SModelS to reevaluate the constraints in a global likelihood analysis. Concretely, we are combining up to 33 analyses from both Run~1 and Run~2 (21 from ATLAS and 12 from CMS) implemented in the \mbox{SModelS}~v3.0.0 database \cite{Altakach:2024jwk} that are sensitive to the electroweak-ino pMSSM scan points. The best (i.e.\ most sensitive) combination is determined for each parameter point separately. Only analyses that can to good approximation be considered as uncorrelated enter the combination.

\begin{figure}[t]\centering
\includegraphics[width=0.48\textwidth]{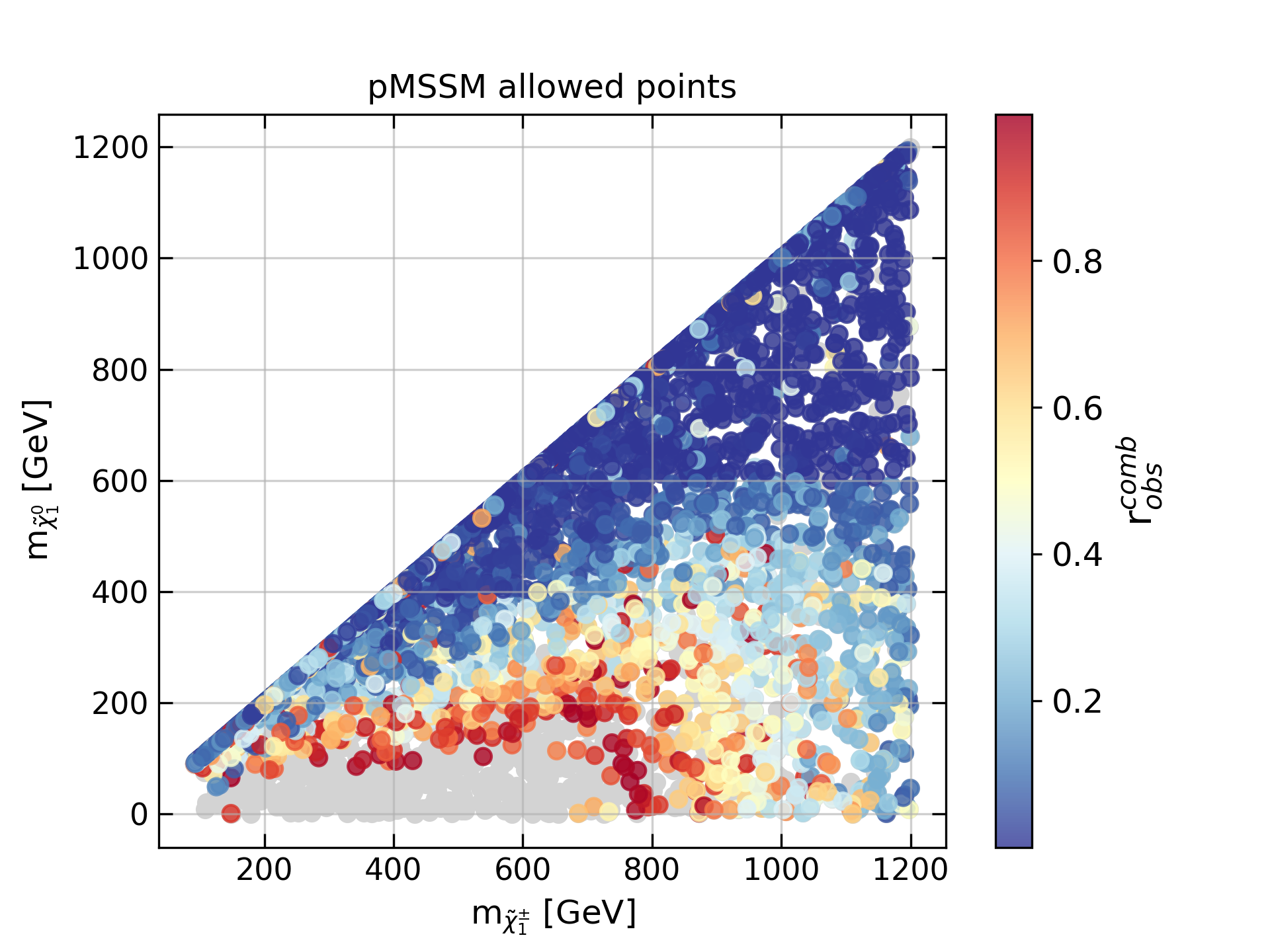}\caption{\small Scatter plot of pMSSM scan points from \cite{ATLAS:2024qmx} that escape exclusion in a global likelihood analysis with SModelS (in colour). Grey points are excluded by the combination of analyses. See text for details.}
\label{fig:ewinoAllowed}
\end{figure}

The result is shown in Fig.~\ref{fig:ewinoAllowed}. The plot shows (in colour) the allowed points, i.e.\ the points not excluded at 95\% confidence level by the global likelihood, in the plane of lightest neutralino ($\tilde\chi^0_1$) versus lightest chargino ($\tilde\chi^\pm_1$) mass. The colour code indicates the combined observed $r$-value, $r^{\rm comb}_{\rm obs}$, with $r$ defined as
the ratio of the predicted fiducial cross section of the signal over the corresponding upper limit.\footnote{Points allowed by the combination but excluded in \cite{ATLAS:2024qmx} by the disappearing tracks analysis with 140/fb, for which only the 36/fb results are available in SModelS, have been removed.}
The grey points have $r^{\rm comb}_{\rm obs}\geq 1$ and are thus excluded by the combination.

As one can see, only a small region of the pMSSM parameter space is definitely excluded, while many scenarios with light charginos and neutralinos remain valid.\footnote{{A similar conclusion was reached in \cite{Barman:2024xlc}, studying the validity of light neutralino thermal dark matter.}} 
The colour code in Fig.~\ref{fig:ewinoAllowed} gives an idea about future prospects:  very roughly, the red/orange/yellow-ish points 
should be within reach with the projected 500~fb$^{-1}$ of data 
at Run~3, while the blue(-ish) points 
may be hard to test (pending the excesses discussed in the next section).  

\begin{figure}[t]\centering
\includegraphics[width=0.48\textwidth]{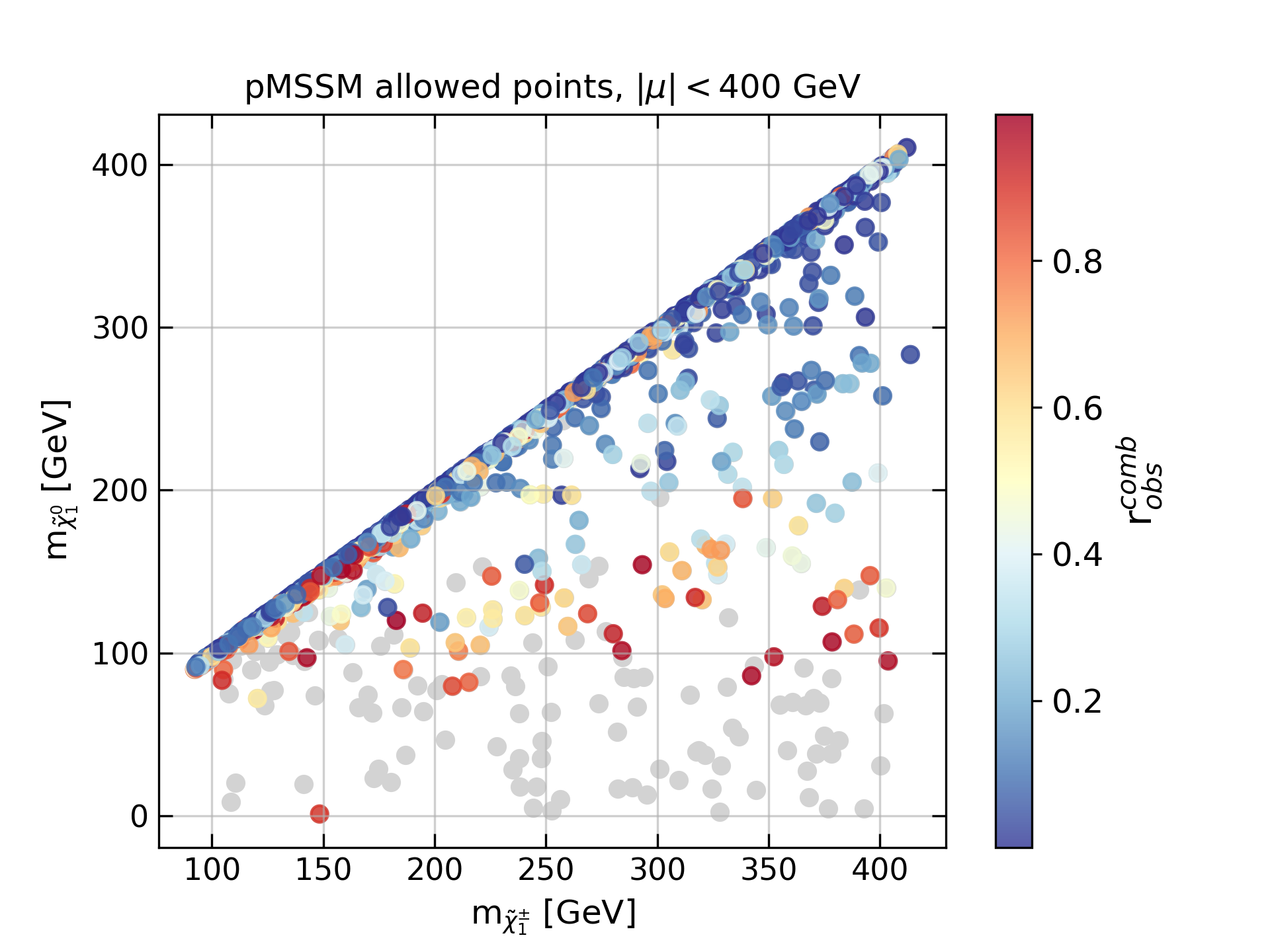}\caption{\small Similar to Fig.~\ref{fig:ewinoAllowed} but requiring $|\mu|<400$~GeV.}
\label{fig:ewinoAllowedSmallmu}
\end{figure}

When requiring $|\mu|<400$~GeV, as preferred by the naturalness argument elaborated on in the previous section, only the region with $m_{\tilde\chi^\pm_1}<400$~GeV remains from  Fig.~\ref{fig:ewinoAllowed}, with the density of points considerably thinned out, see Fig.~\ref{fig:ewinoAllowedSmallmu}.\footnote{The low density of points with $|\mu|<400$~GeV is simply a volume effect from the sampling of a vast parameter space.} The overall picture, however, remains unchanged, with many well-motivated light SUSY scenarios escaping current limits.  
Specifically one finds along the diagonal many higgsino LSP points with very low $r^{\rm comb}_{\rm obs}$, which may be challenging to detect also in future LHC runs; these typically have heavy winos ($|M_2|>1$~TeV). The higgsino LSP points with a promising $r^{\rm comb}_{\rm obs}$ are typically accompanied by winos below about 1~TeV.  There are also points with wino-like LSPs, but with a large enough higgsino or bino admixture to make the $\tilde\chi^\pm_1$ not too long-lived (or even decay promptly). As pointed out in \cite{Carpenter:2023agq}, part of this parameter space could be tested by hadronic mono-boson searches, in particular if the MET cuts are optimised. 
Away from the diagonal, that is for sizeable $\tilde\chi^\pm_1-\tilde\chi^0_1$ mass differences, the spectrum consists of light higgsinos with a bino LSP; again the $r$-values are higher when winos are not too heavy, so they also contribute to the signal. Details will be discussed in \cite{Constantin:2025}.

\section{Hints for light electroweak-inos?}
\label{compressed}

Last but not least, we note that there are tantalising, small but seemingly consistent, excesses in ATLAS and CMS supersymmetry searches involving two or more soft leptons~\cite{ATLAS:2019lng,ATLAS:2021moa,CMS:2021edw,CMS:2024gyw,ATLAS:2025evx}.
The excesses mostly appear in the dilepton ($2\ell$+MET) channel, with a dilepton invariant mass in the range of about $10-20$~GeV, and can be interpreted, e.g., as stemming from light charginos/neutralinos, $pp\to\tilde\chi^\pm_1\tilde\chi^0_2\to W^*\tilde\chi^0_1Z^*\tilde\chi^0_1$, with a compressed mass spectrum, roughly around $\Delta m=m_{\tilde\chi^0_2}-m_{\tilde\chi^0_1}\approx 10-30$~GeV. 
Small excesses are also seen in monojet searches \cite{ATLAS:2021kxv,CMS:2021far} and, as pointed out in \cite{Agin:2023yoq}, may be compatible with the soft lepton ones.  

The simplified models used by ATLAS and CMS for their interpretations, a wino/bino model with $m_{\tilde\chi^0_2}=m_{\tilde\chi^\pm_1}$ 
and a higgsino LSP model with $m_{\tilde\chi^\pm_1}=0.5(m_{\tilde\chi^0_2}+m_{\tilde\chi^0_1})$, are, however, never exactly realised in the MSSM. In the wino/bino case, for the small mass splittings relevant here, there will always be some wino-bino mixing. Moreover, as pointed out in \cite{Baum:2023inl,Chakraborti:2024pdn}, $\tilde\chi^0_2\to \tilde\chi^0_1\gamma$ decays may be relevant and weaken the leptonic signal. In the 
higgsino LSP case, on the other hand, a non-negligible gaugino admixture is needed to make the mass splitting large enough, as pure higgsinos have loop-suppressed mass splittings below about 1~GeV.

Realistic MSSM scenarios for explaining the observed excesses, in part including also dark matter (DM) constraints, were considered in  \cite{Agin:2023yoq,Chakraborti:2024pdn,Agin:2024yfs,Martin:2024pxx}. 
Reference~\cite{Chakraborti:2024pdn} concluded that wino/bino DM in the MSSM performs better than the higgsino scenario (for $\mu>0$) and for       
$m_{\tilde\chi^0_2} \approx m_{\tilde\chi^\pm_1} \le 250$~GeV and $\Delta m = m_{\tilde\chi^0_2}-m_{\tilde\chi^0_1} \approx 25$~GeV, has roughly the right cross section to produce the observed excesses. 
Reference \cite{Martin:2024pxx} considered either sign of $\mu$ and concluded that $\mu<0$ and low $\tan\beta$ (with decoupled heavy Higgs bosons) allows one to evade DM direct detection limits, rendering light higgsinos a viable option. 
In \cite{Agin:2024yfs} the complementarity of the soft lepton excesses with other excesses in LHC monojet searches was  discussed for the MSSM and next-to-MSSM, as well as for two non-SUSY scenarios (a vector-like lepton model and a type-II see-saw model); they also identified the wino/bino scenario in the MSSM as ``the best option currently on the table'', while the non-SUSY interpretations fit the excesses less well.

In any case, more data is needed to assess whether the current small excesses are indeed hints of new physics or mere fluctuations. Run~3 should clarify the situation.

\section{Summary and conclusions}

Has the LHC ruled out supersymmetry? The answer is no! \\
As we have shown, interesting and well-motivated scenarios with light superpartners remain viable even in the simplest incarnation of SUSY, the MSSM. Discovery prospects at the LHC remain good, and there are even possible hints for light electroweak-inos in the current data. 
However, since High-Scale SUSY is also a viable possibility, no currently feasible collider can definitely exclude SUSY as a principle of nature. 

On the other hand, as discussed in  \cite{Dumont:2013npa,Arbey:2015aca,Arbey:2021jdh}, precise measurements of the Higgs boson offer a valuable and complementary probe of supersymmetry. The presence of additional SUSY particles can modify the couplings of the SM-like Higgs, thereby altering its production cross sections and decay branching ratios. As a result, precision Higgs measurements can impose meaningful constraints on the SUSY parameter space, even in scenarios where direct searches are not effective.

All in all, the hunt for SUSY remains exciting.

\section*{Acknowledgements}
We thank the editor for the invitation to write this review.  

\bibliographystyle{elsarticle-num} 
\bibliography{references}

\end{document}